\documentclass[twocolumn,aps,showpacs,prb,amsmath,amssymb,floatfix,superscriptaddress]{revtex4}
\usepackage{color}
\usepackage{amsmath}
\usepackage{amsfonts}
\usepackage{amssymb}
\usepackage{graphicx}
\usepackage{txfonts}
\begin{document}
\title{Universal R\'enyi mutual information in classical systems: the case of kagome ice}
\author{Armin Rahmani}
\affiliation{
Theoretical Division, T-4 and CNLS, Los Alamos National Laboratory, Los Alamos, NM 87545, USA} 
\author{Gia-Wei Chern}

\affiliation{
Theoretical Division, T-4 and CNLS, Los Alamos National Laboratory, Los Alamos, NM 87545, USA}  

\begin{abstract}

We study the R\'enyi mutual information of classical systems characterized by a transfer matrix. We first establish a general relationship between the R\'enyi mutual information of such classical mixtures of configuration states, and the R\'enyi entropy of a corresponding Rokhsar-Kivelson--type quantum superposition. We then focus on chiral and nonchiral kagome-ice systems, classical spin liquids on the kagome lattice, which respectively have critical and short-range correlations. Through a mapping of the chiral kagome ice to the quantum Liftshitz critical field theory, we predict a universal subleading term in the R\'enyi mutual information of this classical spin liquid, which can be realized in the pyrochlore spin ice in a magnetic field. We verify our prediction with direct numerical transfer-matrix computations, and further demonstrate that the nonchiral kagome ice (and the corresponding quantum Rokhsar-Kivelson superposition) is a topologically trivial phase. Finally, we argue that the universal term in the mutual information of the chiral kagome ice is fragile against the presence of defects.

\end {abstract}

\maketitle

\section{introduction}\label{sec:intro}
Universal subleading terms in the entanglement entropy encode valuable information about quantum systems. In gapped topological phases, such as the $Z_2$ spin-liquid ground state of the Kitaev toric-code model,~\cite{Kitaev2003} there is a subleading correction to the area law,~\cite{Hamma2005} known as the topological entanglement entropy, which encodes the total quantum dimension of the system.~\cite{Kitaev2006, Levin2006} 
The ground state of the toric-code model can be viewed as an equal-weight superposition of loops. 
A very similar universal correction can appear in critical wave functions such as the ground state of the Rokhsar-Kivelson (RK)~\cite{Rokhsar1988} quantum dimer models; the so-called RK state is again an equal-weight superposition of dimer coverings.~\cite{Hsu2009,Stephan2009,Hsu2010, Oshikawa2010}
In this latter case, the universal term has a different interpretation: it encodes information about the stiffness constant of the underlying critical field theory, and therefore the exponent of critical correlations.~\cite{Oshikawa2010}

\begin{figure}
\includegraphics[width =0.9\columnwidth]{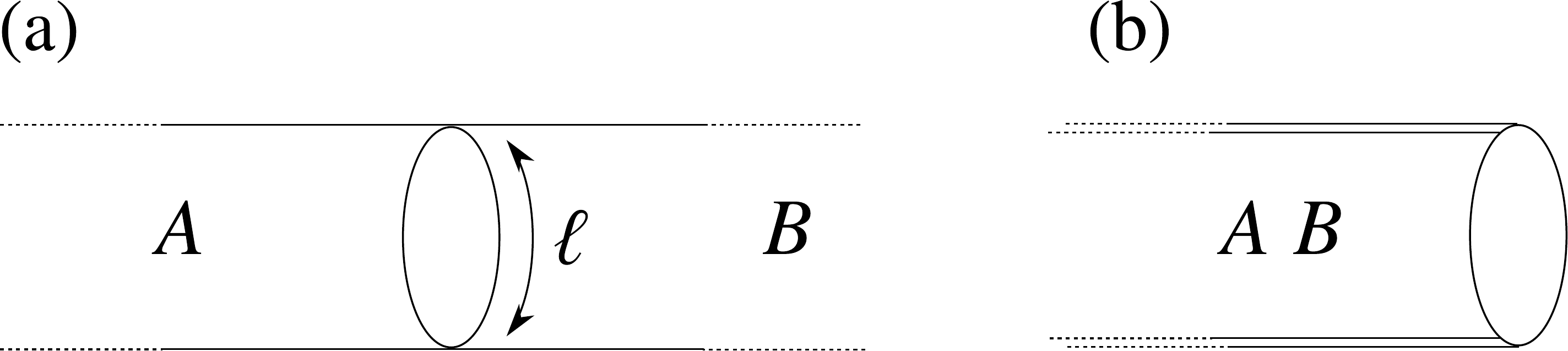}
 \caption{(a) A convenient partitioning for studying universal entanglement is obtained by dividing an infinite cylinder of circumference $\ell$ into two pieces. (b) Folding the $B$ subsystem onto the $A$ half allows for the application of boundary conformal field theory. }
 \label{fig:1}
\end{figure}

The concept of topological order in classical systems has attracted considerable interest recently.~\cite{Castelnovo2007,Henley2011,Hastings2011,Macdonald2011,Jaubert2013,Lamberty2012} In particular, by using the notion of mutual information (a generalization of the entanglement entropy to mixed states), it was shown in Ref.~\onlinecite{Castelnovo2007} that the topological order can survive decoherence and appear in \textit{classical} systems. Moreover, it can be detected through subleading corrections to the mutual information. It was argued in Ref.~\onlinecite{Castelnovo2007} that under fairly general conditions, the classical von Neumann mutual information is half of the corresponding quantum coherent case. Primarily focusing on critical systems, in this work we examine the universal terms in the \textit{R\'enyi} mutual information of classical (thermal) mixture of configuration states, characterized by a transfer matrix, through a direct combinatorial approach. We obtain a general relationship between $I_{AB}^c(n)$, the R\'enyi mutual information of the thermal mixture (where $n$ is the R\'enyi index), and $I_{AB}^q(n)$, of the corresponding quantum superposition:
\begin{equation}\label{eq:relationship}
I_{AB}^c(n)={1 \over 2}I_{AB}^q\left({n+1 \over 2}\right).
\end{equation}
The expression above is valid for a long cylinder divided into two subsystems $A$ and $B$ is in Fig.~\ref{fig:1}(a), and consequently holds, generically, for the universal correction to the area law. This expression generalizes the finding of Ref.~\onlinecite{Castelnovo2007} to an arbitrary R\'enyi index and a wider class of systems. R\'eny mutual information has been also applied recently to studying classical phase transitions.~\cite{Wilms2011, Iaconis2013}

A canonical example of classical thermal mixtures described above is provided by spin-ice systems, which are massively degenerate magnetically disordered manifolds of spin configurations satisfying local constraints broadly referred to as ice rules.~\cite{Bramwell2001} These systems can be thought of as the classical analogs of quantum spin liquids. The entanglement properties of quantum spin liquids (in particular their topological entanglement entropy) have been the subject of numerous recent studies.~\cite{Isakov2011,Zhang2011,Stephan2012,Jiang2012,Poilblanc2012,Grover2013} Despite the prevalence of spin-ice systems in nature, and the connection to quantum spin liquids, the entanglement properties of spin-ice systems are largely unexplored.  Here we focus on two spin-ice systems on the kagome lattice, known respectively as chiral and nonchiral kagome ice. The local constraints, i.e., ice rules, characterizing these spin-ice manifolds allow for a transfer-matrix formulation of the problem, which in turn makes the general relationship~\eqref{eq:relationship} applicable. By using a mapping of the chiral kagome ice to dimers on the honeycomb lattice, we relate the universal correction to the R\'enyi mutual information of this classical spin liquid to universal terms in the R\'enyi entanglement entropy of the Liftshitz quantum critical point,~\cite{Stephan2009,Hsu2010, Oshikawa2010} and find
\begin{equation}\label{eq:kagome}
\gamma_{AB}^c(n)={1 \over 2(1-n)}\ln \left({n+1\over 2}\right),
\end{equation}
where $n$ is the R\'enyi index. As for the nonchiral kagome ice, through a numerical transfer-matrix calculation, we explicitly demonstrate that $\gamma_{AB}^c(n)$ vanishes.

Chiral kagome ice provides an exactly solvable experimentally realized example, where subleading terms, analogous to the topological entanglement entropy, appear in the absence of of quantum coherence or quantum fluctuations. The additional constraints in chiral kagome ice (with respect to the nonchiral manifold) result in critical spin correlations, as well as aforementioned subleading term  $\gamma_{AB}^c(n)$. In contrast, the nonchiral kagome ice has a finite correlation length, extremely weak dependence of the R\'enyi mutual information on the R\'enyi index (which likely vanishes in the thermodynamic limit), and a vanishing topological R\'enyi mutual information. Interestingly, the RK quantum counterpart of the nonchiral kagome ice also has a vanishing topological entanglement entropy. 

 Moreover, we study a continuous interpolation between the chiral and nonchiral kagome-ice systems, and argue that the universal terms~\eqref{eq:kagome} is a special property of the critical point, i.e., it is fragile against the presence of defects. A particular interpolation is obtained by considering pyrochlore spin ice in a [111] magnetic field: the chiral (nonchiral) kagome ice is then realized at temperature $T=0$ ($T=\infty$). We argue that in the thermodynamic limit, the subleading term in the mutual information vanishes throughout the noncritical phase: $\gamma_{AB}^c(n)=0,\quad T>0$, and jumps to the value given in Eq.~\eqref{eq:kagome} only at the critical point.
Finally, our general formulation in terms of a transfer matrix provides a tool for studying the entanglement properties of more complex classical systems such as the three-dimensional pyrochlore spin ice, as well as their RK-type quantum counterparts.

The outline of our paper is as follows. In Sec.~\ref{sec:setup}, we present the general setup of the problem, and derive Eq.~\eqref{eq:relationship}. We describe in Sec.~\ref{sec:kagome}  the chiral and nonchiral kagome-ice systems, and discuss the relationship between the chiral manifold and the Liftshitz quantum critical point. We predict the universal subleading correction~\eqref{eq:kagome} using established results on the Liftshitz quantum critical point,~\cite{Stephan2009,Hsu2010, Oshikawa2010} and verify our prediction with direct numerical calculations. We also explicitly verify the vanishing of the topological R\'enyi mutual information in nonchiral kagome ice, and argue that the universal term~\eqref{eq:kagome} is fragile against thermal defects. We close the paper in Sec.~\ref{sec:conc} with a brief discussion.

\section{general setup}\label{sec:setup}
\subsection{Preliminaries}
The R\'enyi entropy $S_{X}(n)$ of a subsystem $X$ is a measure of entanglement with desired properties such as additivity and continuity:
\begin{equation}\label{eq:renyi}
S_{X}(n)={1 \over 1-n} \ln\left[ {\rm tr}\left(\rho_X^n\right)\right].
\end{equation}
Here the reduced density matrix $\rho_X$ of the subsystem $X$ is defined through tracing out the degrees of freedom in the rest of the system.
The R\'enyi entropy depends on a parameter $n$ known as the R\'enyi index, and the knowledge of the R\'enyi entropies for all $n$ potentially contains more information than the standard von Neumann entropy, which is given by the $n\rightarrow 1$ limit of the R\'enyi entropy.

In fact, for simply connected subsystems, the knowledge of the R\'enyi entropies (for all indices $n$) can yield the full spectrum of the the reduced density matrix, known as the entanglement spectrum, thus providing a more complete characterization of entanglement properties.~\cite{Li2008, Dong2008} While it has been shown that in a certain class of gapped topological phases, the universal correction to the R\'enyi entropy is independent of $n$,~\cite{Flammia2009} for critical and also several gapped phases,~\cite{Jiang2013} the universal correction generally has a nontrivial dependence on $n$.

The R\'enyi entropy is symmetric for a pure-state wave function, i.e., $S_A(n)=S_B(n)$ for a system comprised of two subsystems $A$ and $B$. On the other hand, for generic mixed states we have $S_A(n)\neq S_B(n)$. A natural generalization of the R\'enyi entropy to mixed states, known as the mutual information $I_{AB}(n)$, remedies this problem through an appropriate symmetrization:
\begin{equation}\label{eq:mutual}
I_{AB}={1 \over 2}(S_A+S_B-S_{A\cup B}),
\end{equation}
where $A\cup B$ represents the whole system, and we have suppressed the explicit dependence on the R\'enyi index $n$ for brevity. Since,  for a pure state, $S_{A\cup B}(n)=0$ for all $n$, the mutual information reduces to the R\'enyi entropy of entanglement in this case (the mutual information is often defined without the factor of $1\over 2$ but this factor is convenient because of the connection with entanglement entropy). The mutual information encodes the total amount of quantum and classical correlations between the two subsystems.~\cite{Groisman2005}

For a system exhibiting area law with a universal order-one correction, we have ~\cite{Wolf2008}
\begin{equation}\label{eq:area}
I_{AB}\sim c \ell+\gamma+\cdots,
\end{equation}
where $c$ is a nonuniversal prefactor, $\ell$ is the $(d-1)$-dimensional ``area'' of the  $d$-dimensional subsystem, $\gamma$ is the universal subleading term, and the ellipsis represents other subleading contributions, which vanish in the limit of $\ell \rightarrow \infty$. Note that in some systems there could be logarithmic corrections to area law for subsystems with sharp corners or complex topology,~\cite{Fradkin2006} but here we only consider subsystems with vanishing logarithmic corrections.

A convenient partitioning for extracting universal contributions to the entanglement entropy [see Eq.~\eqref{eq:area}] is obtained by dividing an infinite cylinder of circumference $\ell$ in half.~\cite{Jiang2012} The subsystems $A$ and $B$ then have simple topology and no sharp corners. We thus expect the area law to take the form of Eq.~\eqref{eq:area}, with the subleading terms indicated by an ellipsis vanishing as $e^{-\ell/\xi}$ for gapped phases (where $\xi$ is the correlation length), and as a power-law of $\ell$ for critical systems. In practice, it is helpful to work with a finite cylinder of length $2h$ (with free or fixed boundary conditions at the two endpoints), and take the thermodynamic limit $h\rightarrow \infty$ at the end of the calculation.   

\subsection{Combinatorial argument}
In order to derive Eq.~\eqref{eq:relationship}, let us now consider a general set of (classical) configurations $\cal C$ satisfying some local constraints (which allows for a transfer-matrix formulation). A Hilbert space corresponding to this manifold can be constructed as in the RK model: for each configuration we construct a state $|{\cal C}\rangle$, and impose the orthogonality condition $\langle {\cal C}|{\cal C}'\rangle=\delta_{ {\cal C} {\cal C}'}$. If the degrees of freedom in a given configuration are classical Ising spins, then each state $|{\cal C}\rangle$ is a direct product of single spin-up and -down states $|\!\uparrow\rangle$ and $|\!\downarrow\rangle$. Assuming that there are $\cal N$ such configurations in a given (finite) system, an equal-weight quantum superposition and an equal-weight classical (thermal) mixture, respectively, have the following density matrices:
\begin{equation}\label{eq:rho}
\rho^q={1 \over {\cal N}}\sum_{{\cal C}{\cal C}'}|{\cal C}\rangle \langle{\cal C}' |,\qquad 
\rho^c={1 \over {\cal N}}\sum_{\cal C}|{\cal C}\rangle \langle{\cal C} |.
\end{equation}

\begin{figure}
\includegraphics[width =\columnwidth]{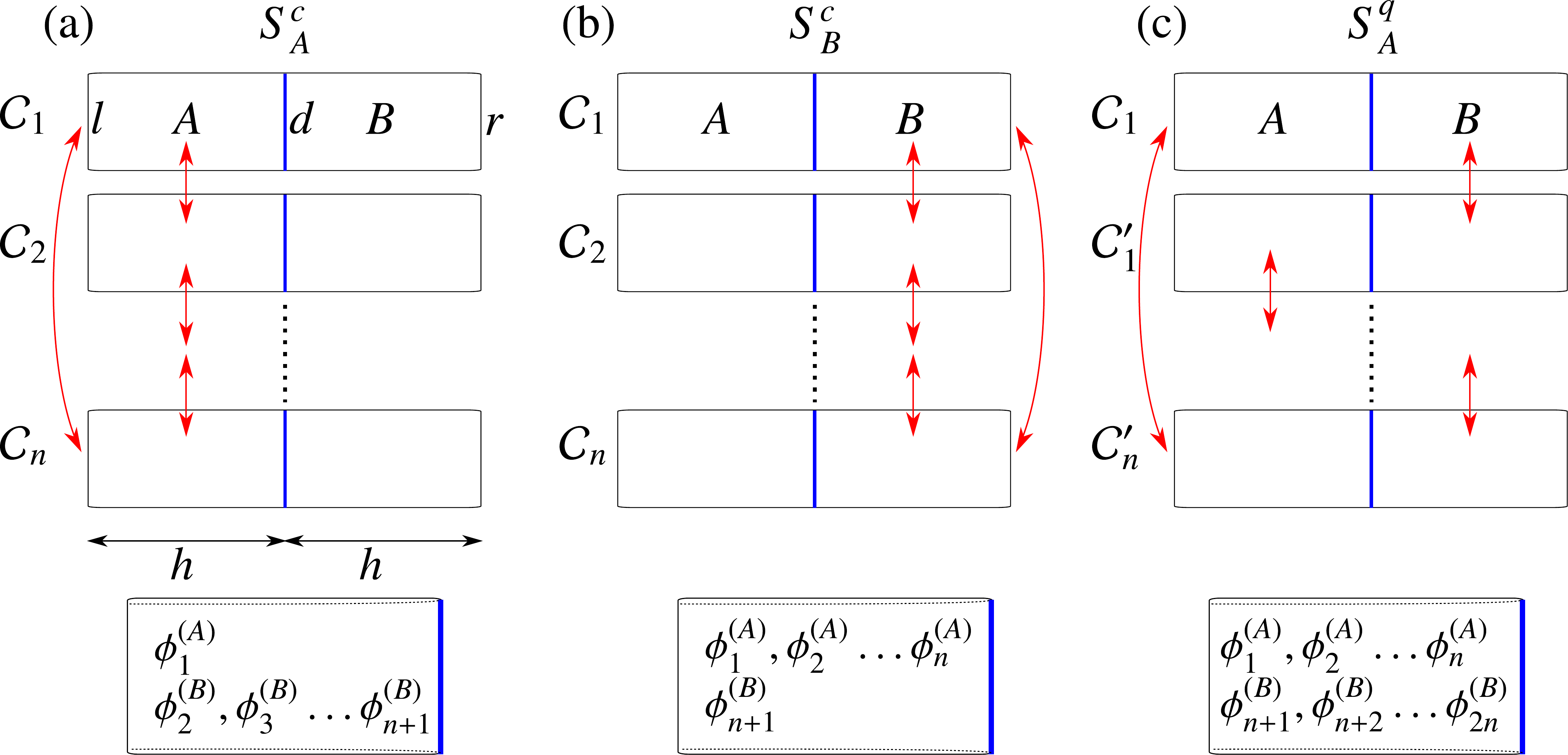}
 \caption{(Color online) (a) To calculate $S_A^c$, we need $n$ replica states $|{\cal C}_i\rangle$, which have the same configurations in $A$ subsystem (top panel). If we fold the $B$ side onto the $A$ side as in Fig.~\ref{fig:1}(b) (bottom panel), we get $n+1$ independent (except at the boundary) sets of configurations (or fields $\phi_i$ in a field-theory representation) living on the half-cylinder. (b) To calculate $S_C^c$, we similarly identify the configurations on the $B$ side. (c) For the quantum case ($S_A^q=S_B^q$), the configurations are identified with a different pattern resulting in $2n$ independent fields in the bulk of the half-cylinder. }
 \label{fig:2}
\end{figure}

We start with the classical density matrix $\rho^c$ in Eq.~\eqref{eq:rho}. Representing the degrees of freedom (say Ising spins) in subsystems $A$ and $B$ with ${\cal A}$ and ${\cal B}$, we can write
\begin{equation}\label{eq:S^c_A}
\begin{split}
S^c_A={1 \over 1-n}\ln\Bigg({1 \over {\cal N}^n}\sum_{{\cal C}_i{\cal A}_i {\cal B}_i}&\langle {\cal A}_1 {\cal B}_1|{\cal C}_1\rangle \langle{\cal C}_1 |{\cal A}_2 {\cal B}_1 \rangle
\langle {\cal A}_2 {\cal B}_2|{\cal C}_2\rangle\dots\\
&\times\langle {\cal A}_n {\cal B}_n|{\cal C}_n\rangle \langle{\cal C}_n |{\cal A}_1 {\cal B}_n \rangle\Bigg),
\end{split}
\end{equation}
where $\{|{\cal A} {\cal B}\rangle\}$ constitutes an orthonormal basis. We further assume that the configuration states $|{\cal C}\rangle$ belong to the set of basis states (which is certainly the case for Ising spins or other local degrees of freedom). We then find that the only contribution to the sum above comes from $|{\cal C}_i\rangle=|{\cal A}_i {\cal B}_i\rangle=|{\cal A}_{i+1} {\cal B}_i\rangle$, which gives ${\cal A}_1={\cal A}_2=\dots {\cal A}_n$. In other words, the $n$ replica configurations ${\cal C}_i$ are ``stitched'' together as shown in Fig.~\ref{fig:2}(a).

 Now if two configurations match in subsystem $A$, they must also match at the boundary between $A$ and $B$. The sum in Eq.~\eqref{eq:S^c_A} above can then be written as a sum over the configurations  $d$ at the boundary as follows:
\begin{equation}\label{eq:S^c_A2}
S^c_A={1 \over 1-n}\ln\Bigg[{1 \over {\cal N}^n}\sum _d N_A(d)N_B^n(d)\Bigg],
\end{equation}
where $N_A(d)$ ($N_B(d)$) denotes the number of configurations in $A$ ($B$) for a fixed configuration $d$ on the boundary [see Fig.~\ref{fig:2}(a)]. The entropy $S^c_B$ can be obtained simply by replacing $A \leftrightarrow B$ in the above expression [see Fig.~\ref{fig:2}(b)]. It is also easy to observe that $S^c_{A\cup B}=\ln {\cal N}$ for all $n$.
A similar treatment for the quantum density matrix $\rho^q$ in Eq.~\eqref{eq:rho} shows that the $2n$ configurations (${\cal C}_i$ and ${\cal C}'_i$) are ``stitched'' together as shown in Fig.~\ref{fig:2}(c) in the quantum case, and we can write
\begin{equation}\label{eq:S^q}
S^q_A=S^q_B={1 \over 1-n}\ln\Bigg[{1 \over {\cal N}^n}\sum _d N_A^n(d)N_B^n(d)\Bigg].
\end{equation}

\subsection{Transfer-matrix approach}
As the configurations $\cal C$ are characterized by local constraints, the combinatorial problem of computing the entropies above can be formulated in terms of a transfer matrix, which has proved useful in the calculation of entanglement entropy in both static~\cite{Stephan2009} and dynamical~\cite{Rahmani2010} cases. Let us break up the system into rows parallel to the boundary between the two subsystems, and indicate the configuration of the degrees of freedom on a row by a lower-case letter (the boundary configuration $d$ indicates the configuration of a row at the interface of the two subsystems). Some care must be taken in identifying the configuration $d$ at the boundary with the configuration of a row, in terms of which the transfer matrix is constructed. Our assumption is that the subsystems $A$ and $B$ either satisfy a symmetry with respect to the boundary, or there are hard local constraints. In the absence of these assumptions, one may need to resort to overlapping the subsystems as in Ref.~\onlinecite{Stephan2009}.

Considering two consecutive rows with configurations $i$ and $j$, we can denote the number of allowed configurations of the degrees of freedom between the two rows by  $T_{ij}$ (if there are no such degrees of freedom $T_{ij}$ is either 0 or 1). We can now represent the row configurations by states and write $T_{ij}=\langle i|T|j\rangle$, where $T$ is the transfer matrix. Representing the states at the left- and right-hand sides of the system, and the boundary between $A$ and $B$ [see Fig.~\ref{fig:2}(a)], respectively, by $|l\rangle$, $|r\rangle$, and $|d\rangle$, we can then write
\begin{equation}
N_A(d)=\sum_l\langle l| T^h|d\rangle, \qquad N_B(d)=\sum_r\langle d| T^h|r\rangle,
\end{equation}
where we have assumed free boundary conditions at the left and right boundaries. Taking the thermodynamic limit, $h\rightarrow \infty$, gives 
\begin{equation}\label{eq:eigen}
\begin{split}
&N_A(d)=\lambda^h\sum_l\langle l| \lambda\rangle\langle \lambda|d\rangle,\qquad
N_B(d)=\lambda^h\sum_r\langle d| \lambda\rangle\langle \lambda|r\rangle,
 \end{split}
\end{equation}
where $\lambda$ is the largest eigenvalue of $T$, and $|\lambda\rangle$ is the corresponding eigenvector. Similarly, we have ${\cal N}=\sum_{lr}\langle l| T^{2h}|r\rangle=\lambda^{2h}\sum_{lr}\langle l| \lambda\rangle\langle \lambda|r\rangle$.

Note that for fixed (Dirichlet) boundary conditions (instead of free), we can use one particular $|r\rangle$ or $|l\rangle$ and omit the summation over $r$ and $l$. Our final result will not change as the summation over the boundary conditions cancels out but we may need to choose a different $|\lambda\rangle$: the largest-eigenvalue eigenstate of $T$, which has a nonvanishing overlap with both fixed $|r\rangle$ and $|l\rangle$.
 Also note that, more generally, the density matrices may not represent equal-weight mixtures (or superpositions in the quantum case). As discussed in the Appendix, however, as long as the configuration-dependent weights are ultralocal, they can be absorbed into the transfer matrix between consecutive rows (in this case the number of configurations with given boundary conditions transforms into a partition function with those boundary conditions).

Putting all these together, and assuming the overlaps between $|\lambda\rangle$ and the boundary states above are real, we then obtain
\begin{equation}
S_A^c={1 \over 1-n}\ln\left[\frac{\lambda^{h(n+1)}\left(\sum_d \langle \lambda|d\rangle^{n+1}\right)}{\lambda^{2hn}\left(\sum_l\langle \lambda|l\rangle\right)^{n-1}}\right].
\end{equation}
For this particular geometry, the expression for $S_B^c$ is given by changing the dummy variable $l$ to $r$ (and yields an identical result). Similarly, we can write $S_{A\cup B}^c=-{1 \over 1-n}\ln\left[\lambda^{2h(n-1)}\left(\sum_l\langle \lambda|l\rangle\right)^{2n-2}\right]$, which leads to the general expression
\begin{equation}\label{eq:I_classical}
I_{AB}^c(n)={1 \over 1-n}\ln\left(\sum_d \langle \lambda|d\rangle^{n+1}\right),
\end{equation}
where the R\'enyi index $n$ is written explicitly. 
The entanglement entropy of the quantum case was studied in Ref.~\onlinecite{Stephan2009} using this transfer matrix method. For completeness, we restate that result in our notation: by inserting Eq.~\eqref{eq:eigen} into Eq.~\eqref{eq:S^q}, we immediately obtain
\begin{equation}
I_{AB}^q(n)={1 \over 1-n}\ln\left(\sum_d \langle \lambda|d\rangle^{2n}\right).
\end{equation}

Notice that the row configurations $d$ form an orthonormal basis and the normalization of the eigenvector $|\lambda\rangle$ implies that $\sum_d \langle \lambda|d\rangle^2=1$. In other words, $p_d\equiv \langle \lambda|d\rangle^2$ is a probability, and the above expression is identical to that of Ref.~\onlinecite{Stephan2009}.
Comparing the two equations above, we finally arrive at the general relationship $I_{AB}^c(n)={1 \over 2}I_{AB}^q\left({n+1 \over 2}\right)$, which is one of the main results of this paper [see Eq.~\eqref{eq:relationship}], and provides a nontrivial generalization of the prediction of Ref.~\onlinecite{Castelnovo2007} to arbitrary $n$. It is also worth mentioning that subtle issues (such as the inter-dependence of the compactification radii of fields) make it difficult to apply the field-theoretical approach of Ref.~\onlinecite{Castelnovo2007} to certain problems such as RK dimer models.~\cite{Hsu2009,Stephan2009,Hsu2010, Oshikawa2010} Despite this, the result of our combinatorial approach, which directly applies to dimer models, is in perfect agreement with Ref.~\onlinecite{Castelnovo2007} for $n\rightarrow 1$.

\section{kagome ice and quantum liftshitz model}\label{sec:kagome}
\subsection{Kagome-ice manifolds}
\begin{figure}
\includegraphics[width =0.9\columnwidth]{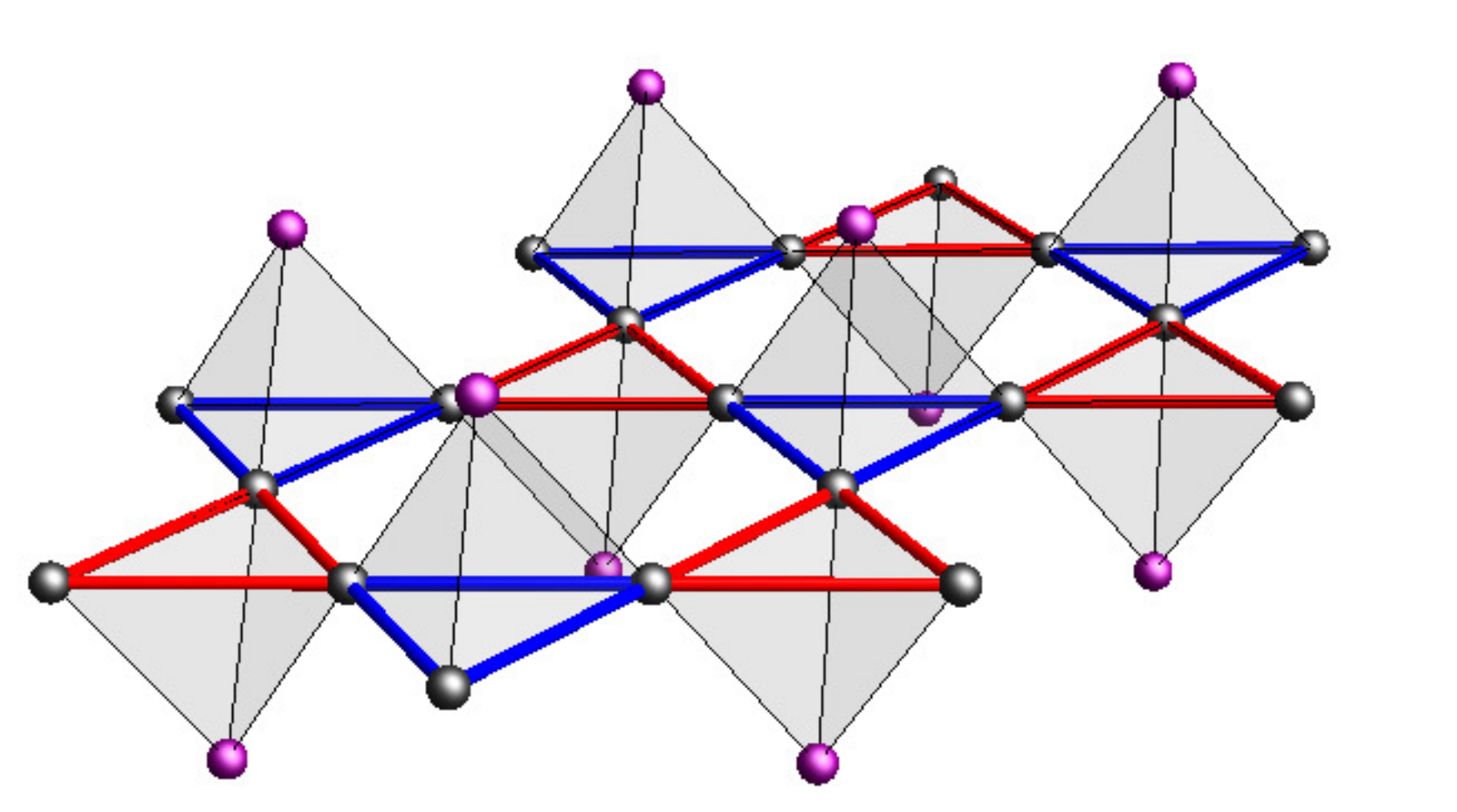}
 \caption{(Color online) A kagome layer in the pyrochlore lattice.}
 \label{fig:3}
\end{figure}
Highly frustrated magnets are canonical examples of classical spin liquids with nontrivial correlations. In particular, the so-called spin-ice materials in which Ising-type moments reside on a three-dimensional network of corner-sharing tetrahedra (the pyrochlore lattice shown in Fig.~\ref{fig:3}) exhibit a critical dipolar-like spin correlations in an emergent Coulomb phase.~\cite{Henley2010} The origin of this critical phase can be traced to the nontrivial local constraints known as the two-in-two-out ice rules that govern the ordering of magnetic moments in individual tetrahedra.~\cite{Bramwell2001}

\begin{figure}
\includegraphics[width =0.9\columnwidth]{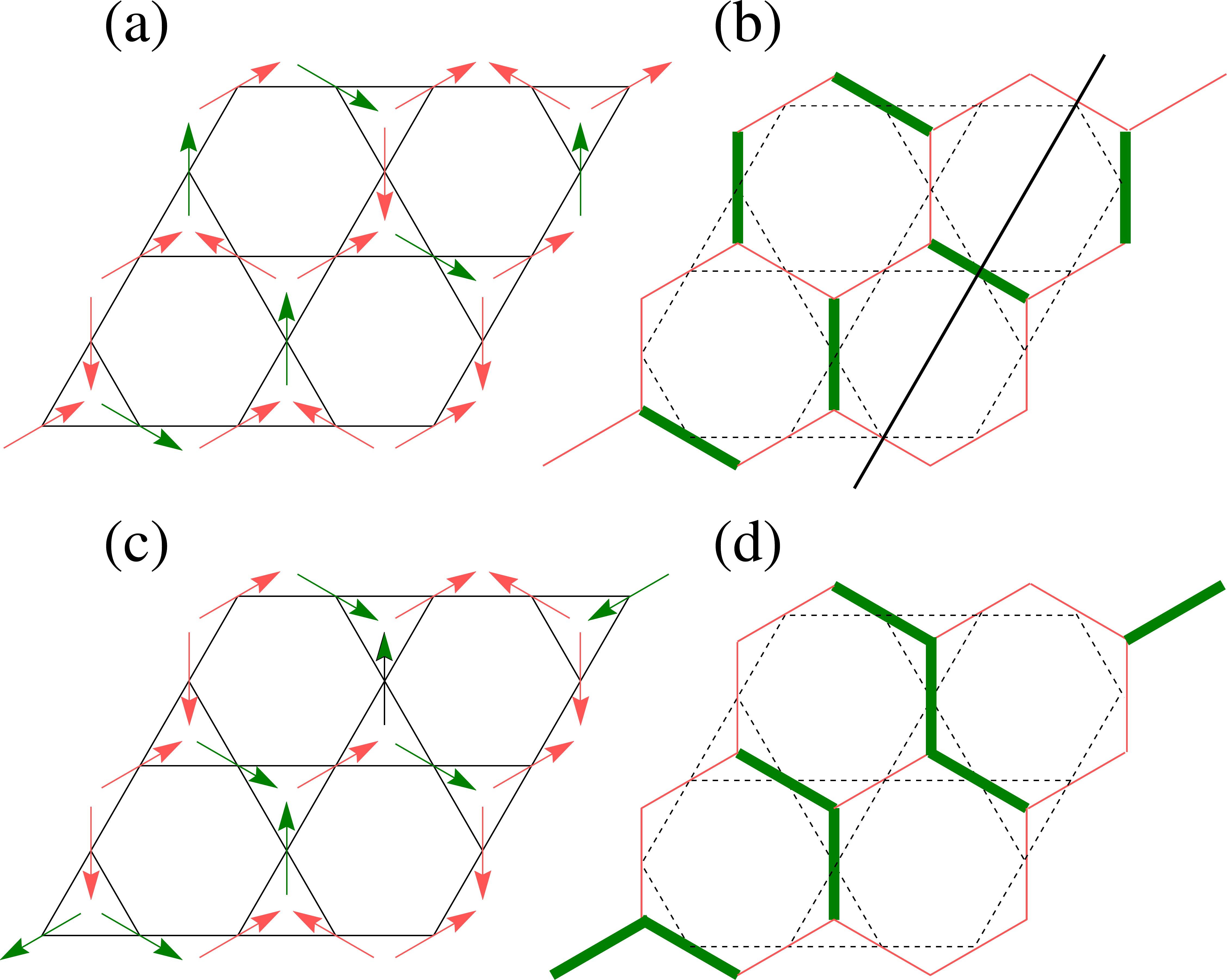}
 \caption{(Color online) (a) A chiral kagome ice configuration. (b) The mapping to dimer coverings. The configurations of the bonds cut across by the solid black line represent a boundary state. (c) A nonchiral kagome-ice configuration. (d) The mapping to dimers with one type of defect. }
 \label{fig:4}
\end{figure}

In this work, we apply the above analysis to two-dimensional analogs of the pyrochlore spin ice: (i) chiral and (ii) nonchiral kagome ice. The kagome lattice shown in Fig.~\ref{fig:4}(a) consists of corner-sharing triangles with two different orientations. The projections of the easy axes on the kagome plane form 120-degree ordering as in Fig.~\ref{fig:4}(a). In chiral kagome ice, every up triangle has a two-in-one-out spin configuration with a net positive magnetic charge, while every down triangle has a one-in-two-out configuration with a net negative charge [see Fig.~\ref{fig:4}(a)]; the entropy density of this ice phase is $0.108\,k_B$ per spin.~\cite{Udagawa2002, Moessner2003} Since the two types of triangles contain opposite magnetic charges, hence breaking the sublattice symmetry, this phase is referred to as the chiral kagome ice. The nonchiral kagome ice is a less constrained manifold: every triangle (whether up or down) is in either the two-in-one-out or the one-in-two-out state [see Fig.~\ref{fig:4}(c)]; the entropy density of this ice phase is $0.501\,k_B$ per spin.~\cite{Wills2002}.

The nonchiral degenerate manifold minimizes the nearest-neighbor spin interaction energy. This ice phase is noncritical with short-range spin-spin correlations. The chiral kagome ice appears in the intermediate plateau regime of the pyrochlore spin-ice in a [111] magnetic field,~\cite{Matsuhira2002,Udagawa2002, Moessner2003} where the spins align themselves with the field on the triangular layers, and, consequently, form the degenerate chiral manifold on the kagome layers (see Fig.~\ref{fig:3}). An artificial version of the kagome ice is also realized in arrays of single-domain ferromagnetic nanoisland arranged in a honeycomb network.~\cite{Tanaka2006,Qi2008} In artificial kagome ice, the chiral ice phase is induced by further-neighbor spin-spin interactions.~\cite{Chern2011,Moller2009} 

Remarkably, spin-spin correlations in chiral kagome ice decay algebraically.~\cite{Moessner2001} The critical correlations of this phase can be understood by mapping the chiral ice configurations to dimer coverings on the dual honeycomb lattice obtained by connecting the centers of the corner-sharing triangles of the kagome lattice~\cite{Moessner2001, Udagawa2002,Moessner2003} [see Fig.~\ref{fig:4}(b)]. Since every triangle has exactly one negative spin (lying on a bond of the honeycomb lattice), each site of the honeycomb lattice is visited by exactly one negative spin, which we identify with a dimer. The dimer-covering problem on honeycomb is exactly solvable using the well-known Pfaffian method for computing the partition function.~\cite {Kasteleyn} Exact calculations showed a power-law dimer-dimer correlation decaying as $1/r^2$.~\cite{Yokoi1986}  In the language of dimers, the nonchiral kagome ice then corresponds to dimer coverings on the honeycomb lattice with one type of defect: a site visited by exactly two dimers [see Fig.~\ref{fig:4}(d)].

\subsection{Chiral kagome ice and Liftshitz quantum model}
In order to obtain the universal correction term to the R\'enyi mutual information for the chiral kagome ice [Eq. \eqref{eq:kagome}], we first consider the R\'enyi entropy of the corresponding quantum RK state and then invoke the classical-quantum correspondence~(\ref{eq:relationship}). Universal entanglement entropy in RK wave functions of dimer coverings is a well-studied problem,~\cite{Hsu2009,Stephan2009,Hsu2010, Oshikawa2010} but the connection to spin-ice systems has not been exploited to study the the entanglement properties of these classical spin liquids. Let us briefly review the field-theoretical approach to the quantum RK case.
The RK wave function $|\Psi\rangle$ is an equal weight superposition of all dimer coverings on the honeycomb lattice, or equivalently an equal weight superposition of all chiral kagome-ice states.  

Dimer models on a bipartite lattice in turn have a height-model representation: scalar heights are defined on every hexagonal plaquette, when a small circular path around a site (which crosses three bonds) goes across an empty bond (dimer), we decrease (increase) the height by $1$ ($2$). The path is directed clockwise for sites on one sublattice and counterclockwise for sites on the other. We then obtain a unique height field modulo the height of one reference plaquette. A periodic structure is imposed on the heights to avoid overcounting configurations.~\cite{Ardonne2004} Upon coarse-graining, the height-model representation then leads to a critical field theory of compactified noninteracting bosons (coarse-grained height field) with effective action:~\cite{Henley1997,Zeng1997}
\begin{equation}
S[\phi]={g\over 4 \pi}\int d^2 x (\partial_\mu \phi)^2, \quad \phi \sim\phi+2\pi R,
\end{equation}
which represents a universality class characterized by the dynamical exponent $z=2$, stiffness $g$, and the compactification radius $R$. By rescaling the bosonic fields $\phi$, we observe that the only free parameter of the theory is $R\sqrt{g}$, which can be computed by comparing the dimer correlation functions obtained from the above continuum quadratic filed theory with exact results based on the Pfaffian method.~\cite{Yokoi1986} 

The projection of the RK wave function to a particular height field configuration and the quantum density matrix [see Eq.~\eqref{eq:rho}] are given by
\begin{equation}\label{eq:wavefunc}
\langle \phi|\Psi\rangle\propto e^{-S[\phi]}, \quad \rho^q=|\Psi\rangle \langle \Psi|.
\end{equation}
The calculation of the entanglement entropy in the cylinder geometry then involves folding half of the cylinder (subsystem $B$) onto the other half. The identification of Fig.~\ref{fig:2}(c) results in $2n$ ($n$ form each subsystem) independent (in the bulk) fields with say Dirichlet boundary condition on the left-hand side of the folded system (the far ends of both $A$ and $B$ subsystems), and a special boundary condition $\phi_1=\phi_2=\dots=\phi_{2n}$ on the right-hand side of the folded system (the boundary between the $A$ and $B$ subsystems).

By explicitly constructing the boundary state corresponding to this boundary condition, it was shown in Ref.~\onlinecite {Oshikawa2010} (see also Refs.~\onlinecite{Stephan2009,Hsu2010}) that the universal R\'enyi entropy should go as ${1 \over 1-n}\ln\left[(\sqrt{2g} R)^{-({N}/2-1)}\sqrt{{N}\over 2}\right]$, where $N$ is the number of independent fields, which in the quantum case equals $N=2n$. Exactly the same argument also holds for the classical case except that, as demonstrated in Figs.~\ref{fig:2}(a) and \ref{fig:2}(b), the number of independent fields is $N=n+1$ ($n$ from one subsystem and $1$ from the other) with the boundary condition $\phi_1=\phi_2=\dots=\phi_{n+1}$ on the right-hand side of the folded system. This is consistent with our Eq.~\eqref{eq:I_classical} found through an explicit combinatorial approach. The actual stiffness of the dimer model corresponds to $\sqrt{2g} R=1$ (see, e.g., Ref.~\onlinecite{Stephan2009}), which immediately leads to our prediction [Eq.~\eqref{eq:kagome}] of the universal mutual information in chiral kagome ice. Note that a phase transition has been predicted in quantum dimer models as a function of the R\'enyi index $n$ with a critical R\'enyi index $n_{\rm c}=9$.~\cite{Stephan2011} Thus, our results for the classical case should only hold for $n<2\times 9-1=17$.

We also explicitly verified Eq.~\eqref{eq:kagome} with the transfer-matrix approach, i.e., by using Eq.~\eqref{eq:I_classical}. We assume that the boundary between the $A$ and $B$ subsystems cuts across $\ell$ bonds of the honeycomb lattice [see Fig.~\ref{fig:4}(b)]. The transfer matrix is then a $(2^\ell \times 2^\ell)$-dimensional matrix, which can be constructed using the local constraints (counting the number of dimer coverings by transfer matrices was pioneered by Lieb.~\cite{Lieb}). The transfer matrix on the honeycomb lattice has a block-diagonal structure: if the transfer matrix has a nonzero matrix element $\langle i|T|j\rangle$, the states $|i\rangle $ and $|j\rangle$ must have the same number of dimers. A mapping from dimers to fermions yields elegant analytical results for the transfer matrix both on the square~\cite{Alet2006} and the honeycomb lattices.~\cite{Stephan2009} 

\begin{figure}
\includegraphics[width =0.9\columnwidth]{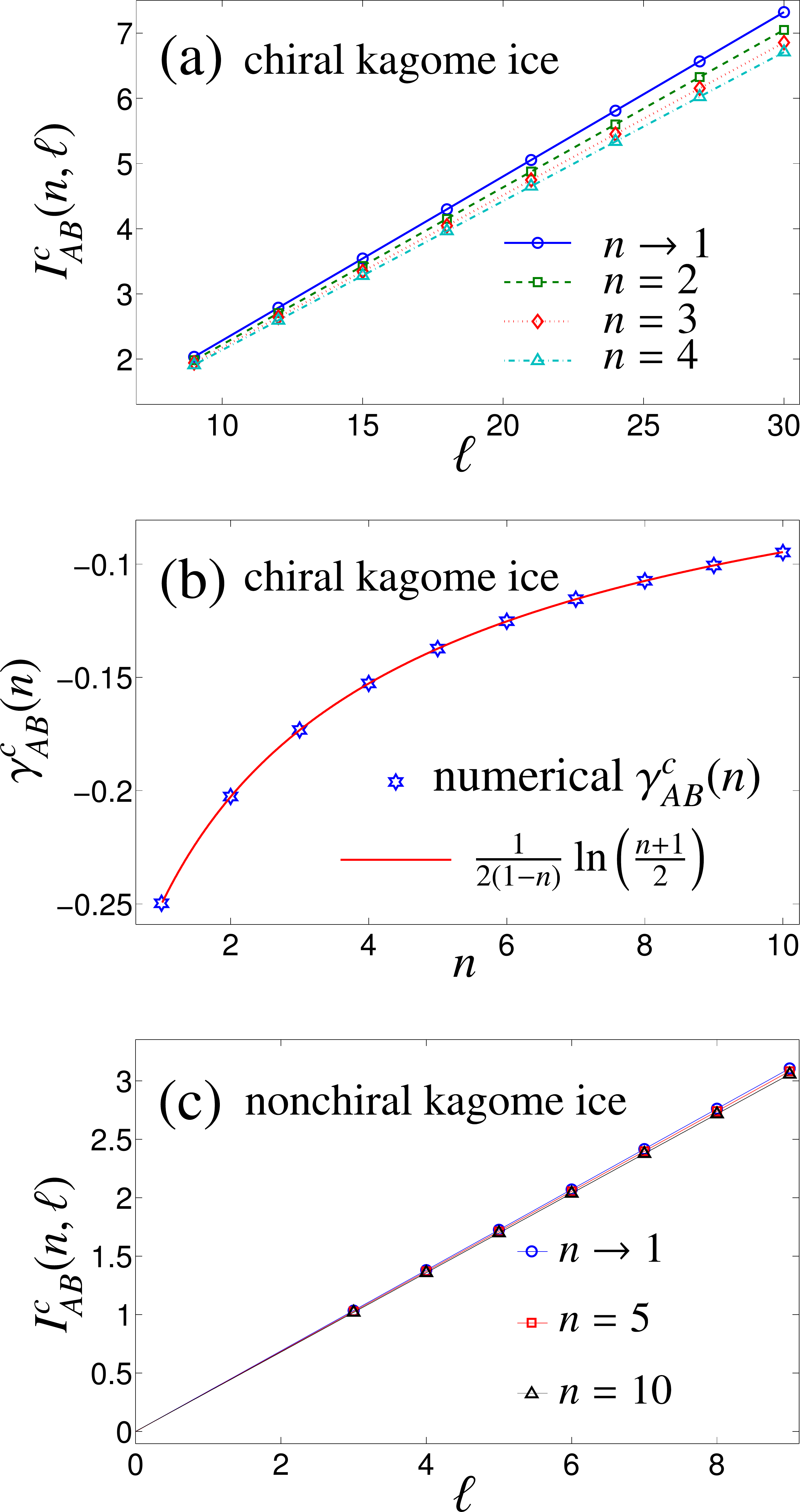}
 \caption{(Color online) (a) Numerically computed mutual information using Eqs.~\eqref{eq:overlap} and \eqref{eq:I_classical} exhibits the area law.  (b) Fitting the data in panel (a) gives the universal piece $\gamma$ (blue stars), which are in perfect agreement with the analytical prediction Eq.~\eqref{eq:kagome} (red line). (c) For the nonchiral kagome ice, the mutual information has a very weak dependence on the R\'enyi index $n$, and the linear fit to data gives $\gamma^c_{AB}\approx 0$.  }
 \label{fig:5}
\end{figure}

In the case of free boundary conditions at infinity, we need overlaps $\langle \lambda|i\rangle$ for the eigenstate with the largest eigenvalue and all of the $2^\ell$ row states. As the transfer matrix is block diagonal, we can consider only the states in the block with the largest eigenvalue. It was shown in Ref.~\onlinecite{Stephan2009} that the number of dimers in this block is $\ell/3$. Denoting the positions of dimers in state $|i\rangle$ as $\alpha_i$ with $0\leqslant \alpha_1<\alpha_2<\dots< \alpha_{\ell/3} \leqslant \ell-1$, the mapping to fermions allows us to compute $|\langle \lambda|i\rangle|^2$ as a Vandermonde determinant:~\cite{Stephan2009}
\begin{equation}\label{eq:overlap}
|\langle \lambda|i\rangle|^2={1 \over \ell^{\ell/3}}\prod_{1\leqslant j<j'\leqslant \ell/3} 4 \sin^2\left[{\pi \over \ell}(\alpha_j-\alpha_{j'})\right].
\end{equation}
The above expression provides an efficient way to compute the R\'enyi entropy for large $\ell$ (here we went up to $\ell=30$). We have also checked the result by a brute-force method (numerical construction of the block-diagonal transfer matrix and direct exact diagonalization) for $\ell$ up to $18$. 
The results are shown in Fig.~\ref{fig:5}(a) for $n=1\dots 4$. As expected, for each $n$, the mutual information $I^c_{AB}(n,\ell)$ satisfies the area law, i.e., scales as $\propto \ell $ for large $\ell$.  To extract the universal subleading term $\gamma$, we fit these data to $I=c\ell+\gamma+c'/\ell$, and obtained $\gamma$ as a function of $n$. The agreement with Eq.~\eqref{eq:kagome} is excellent as seen in Fig.~\ref{fig:5}(b).

\subsection{Nonchiral kagome ice and defects}
It is also illuminating to consider the case of nonchiral kagome ice, which has less constrained ice rules than the chiral manifold. The analogous quantum wave function then has a finite correlation length. The subleading correction to the area law in this case can then be interpreted as topological entanglement entropy. We find through numerical transfer-matrix calculations that the mutual information satisfies the area law, and a fit to $c\ell+\gamma$ gives $\gamma$ of order $10^{-4}$, i.e., the nonchiral kagome ice (and the corresponding quantum RK wave function) is topologically trivial. This provides an interesting example where a quantum wave function, with a superficial resemblance to a quantum spin liquid, is actually a topologically trivial paramagnet. Notice that the transfer matrix does not have a block-diagonal structure in the nonchiral case, and we do not have an exact analytical solution for the overlaps $\langle\lambda |i\rangle$ so we are limited to smaller systems in the numerics. However, we do not need to go to larger systems because the finite-size corrections are exponentially small. The universal terms can then be easily extracted from a simple linear fit for very small system sizes.

One comment is in order before proceeding. In the transfer-matrix calculation above, we used a $2^{\ell} \times 2^{\ell}$ transfer matrix for the configuration of a row shown in Fig.~\ref{fig:4}(b). The subsystems are constructed such that all the degrees of freedom between the the last row of $A$ and the first row of $B$ belong to one of the two subsystems, say $A$. In the chiral kagome-ice case, the first row of $B$ is uniquely determined by these degrees of freedom and the last row of $A$. In the nonchiral case, however, the constraints are less rigid and identifying the boundary configuration $d$ [see Eq. ~\eqref{eq:S^c_A2}] with a row as in Fig.~\ref{fig:4}(b) is rather subtle. We have checked  (for smaller systems), however, that using a $2^{3\ell}\times 2^{3\ell}$ transfer matrix (so that there are no degrees of freedom between two consecutive rows and the two subsystems are symmetric with respect to the boundary) leads to the same result. Such microscopic considerations do not appear in the continuum field-theoretic approach, and seem unlikely to play an important role in general.

An interesting question is the behavior of $\gamma$ when we interpolate between these two manifolds. If we consider the pyrochlore spin ice in a magnetic field $H$ in the [111] direction, we have a density matrix that only depends on the ratio $H/T$ at temperature $T$. In terms of the dimer model, there is an energy cost proportional to $H$ for any defect (two dimers touching on one site). At zero temperature (and $H>0$), $H/T$ diverges and no defects are allowed, which results in the critical chiral kagome-ice manifold. On the other hand, the nonchiral ice corresponds to $H/T=0$, where there is no cost for such defects. At low temperatures, any infinitesimal finite temperature destroys the critical phase, resulting in a finite density of defects and a correlation length that is exponentially large in $H/T$.~\cite{Moessner2003} This instability follows from the fact that in the height representation, a defect is like a vortex (the compactified height field winds once when going around such defect) with an operator proportional to $\cos \left(\sqrt{2} \theta\right)$, where $\theta$ is the dual field to the height field $\phi$, which is a relevant (in renormalization-group sense) perturbation.~\cite{Otsuka2011}. If we compute $\gamma$ in a finite system (smaller than the correlation length) at small finite temperatures, our results changes continuously from the $T=0$ results [Eq.~\eqref{eq:kagome}]. However, if we take the thermodynamic limit first, we do not expect $\gamma$ to change within a noncritical phase as $\xi/\ell$ vanishes if we take the limit of $\ell\rightarrow \infty$ before any other limit. We are thus lead to conjecture that the universal term~\eqref{eq:kagome} is fragile against any density of defects in the thermodynamic limit.

\begin{figure}
\includegraphics[width =0.9\columnwidth]{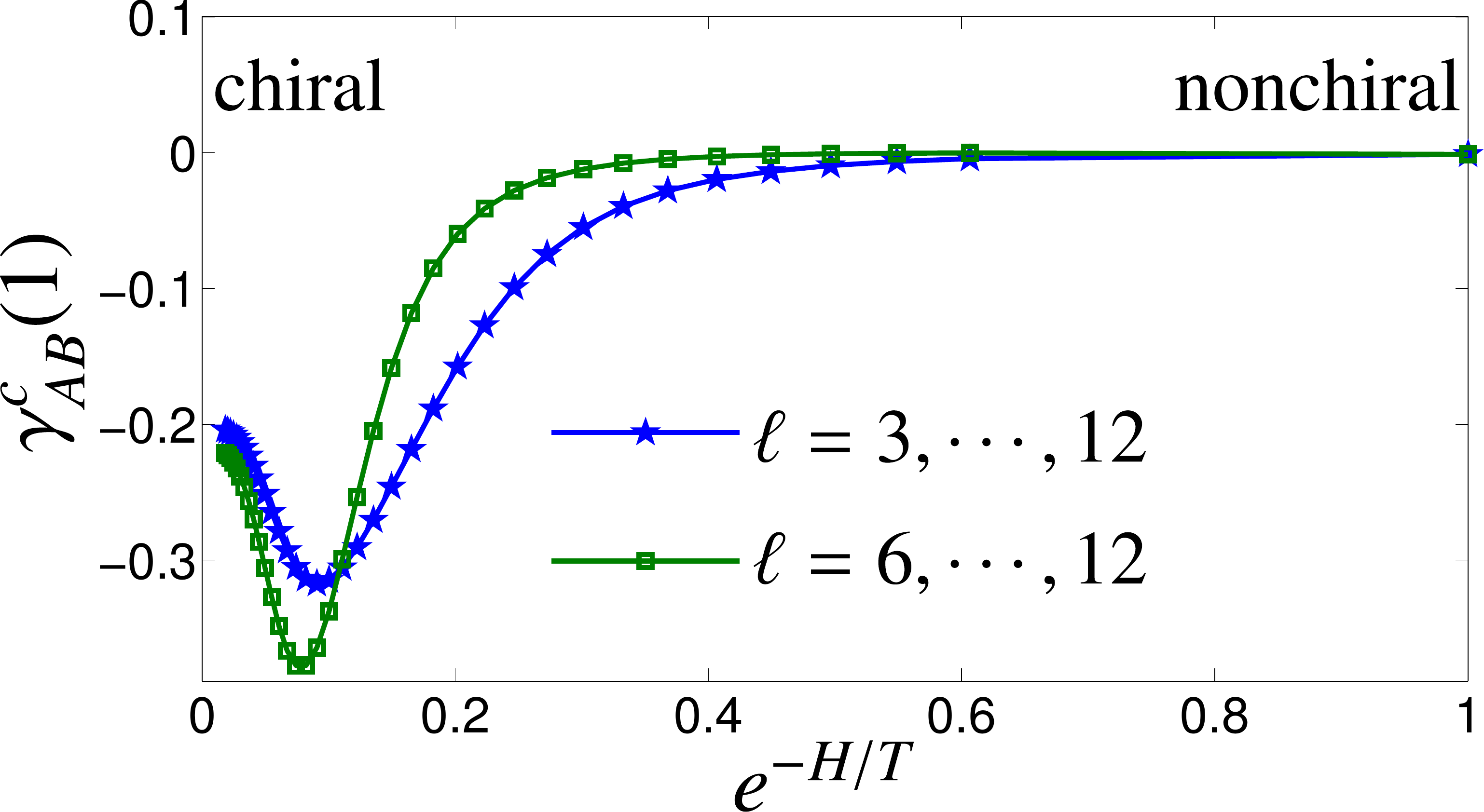}
 \caption{(Color online) The von Neuman mutual information as a function of $e^{-H/T}$, where $H$ is the energy cost of one defect. By excluding smaller systems in the fitting, we observe that the dramatic change away from the $\gamma=0$ plateau is shifted toward lower temperatures.}
 \label{fig:6}
\end{figure}

For finite $H/T$, where we have a small correlation length, we have performed numerical transfer-matrix calculations, and verified the presence of a $\gamma=0$ plateau as we move away from the nonchiral kagome ice corresponding to $H/T=0$. As the correlation length keeps increasing with $H/T$, however, the calculation becomes less reliable in the vicinity of the  the chiral phase (note that we can not use the exact solution of the chiral kagome ice, which allowed for large system sizes in the presence of defects). Nevertheless, by excluding smaller system sizes in fitting the data, we observe a very suggestive trend shown in Fig.~\ref{fig:6}: the dramatic change from $\gamma$ is shifted toward smaller temperatures, supporting our conjecture of fragility. Similar behavior has been proposed in the topological Kitaev toric code model.~\cite{Castelnovo2007b,Nussinov2009a, Nussinov2009b}

\section{discussion}\label{sec:conc}

Motivated by the predictions of Ref.~\onlinecite{Castelnovo2007} on the appearance of topological entanglement entropy in classical systems, which arise through the decoherence of topologically ordered quantum wave functions, we obtained a generalized relationship [Eq.~\eqref{eq:relationship}] between the R\'enyi mutual information of generic classical systems (which can be described by a transfer matrix), and their quantum-coherent RK-type counterparts. The classical systems above can be similarly viewed as decohered versions of these RK quantum wave functions.

We examined the universal entanglement properties of two spin-ice manifolds (chiral and nonchiral) on the kagome lattice, and found a nonvanishing universal subleading term in the mutual information of the chiral state (a highly constrained classical spin liquid with critical correlations). The experimentally relevant chiral kagome ice, which appears in kagome layers of pyrochlore spin ice in the presence of a [111]-direction magnetic field, readily maps onto a dimer model on the honeycomb lattice. The RK quantum dimer model, and its well-known corresponding quantum Liftshitz universality class, thus describe the quantum-coherent counterpart of the chiral kagome ice. This mapping, and the general relationship~\eqref{eq:relationship} thus leads to our prediction of the presence of a universal subleading term in chiral kagome ice, which bears a striking resemblance to the topological entanglement entropy in gapped topologically ordered phases.

The universal mutual information in this critical classical spin liquid, has a nontrivial dependence on the R\'enyi index. Moreover, we verified that for the short-range correlated nonchiral kagome ice, the dependence of the mutual information on the  R\'enyi index is extremely weak (likely only in the subleading terms), and the topological mutual information vanishes. The quantum counterpart of the nonchiral kagome ice provides an interesting example of a trivial paramagnet, which, on the surface, resembles a topological quantum spin liquid. The topologically trivial phase extends even when $Z_2$ chirality is explicitly broken by a magnetic fields. In other words, in the thermodynamic limit, the universal term obtained at the critical point is fragile to any infinitesimal density of defects.

\acknowledgements{We thank C. Batista, C. Castelnovo, C. Chamon, E. Fradkin, B. Hsu, and G. Misguich for helpful discussions and comments. We are specially grateful to I. Martin for collaboration in the early stages of this work and for several helpful suggestions. This work was supported by U.S. DOE under the LANL/LDRD program.}

\appendix
\section{DENSITY MATRICES WITH NONEQUAL WEIGHTS}
In this appendix, we outline how the derivation of Eq.~\eqref{eq:relationship} changes when the density matrices have nonequal (but ultralocal) weights. The quantum and classical density matrices can be written as
\begin{equation}\label{eq:rho2}
\rho^q={1 \over {\cal Z}(\beta)}\sum_{{\cal C}{\cal C}'}e^{-{\beta\over 2}(E_{\cal C}+E_{{\cal C}'})}|{\cal C}\rangle \langle{\cal C}' |,\quad 
\rho^c={1 \over {\cal Z}(\beta)}\sum_{\cal C}e^{-\beta E_{\cal C}}|{\cal C}\rangle \langle{\cal C} |,
\end{equation}
where the partition function is given by ${\cal Z}(\beta)=\sum_{\cal C}e^{-\beta E_{\cal C}}$. Computing $S^c_A$ gives
\begin{equation}\label{eq:ScA2}
S^c_A={1\over 1-n} \ln\left[ {1\over {\cal Z}^n(\beta)}\sum _{{\cal A}{\cal B}_i} e^{-\beta\left(E_{{\cal A}{\cal B}_1}+E_{{\cal A}{\cal B}_2}+\cdots E_{{\cal A}{\cal B}_n}\right)}\right].
\end{equation}
Once again all configurations ${\cal A}{\cal B}_i$ must match at the boundary between $A$ and $B$ represented by the row configuration $d$ so we can write the generalized version of Eq.~\eqref{eq:S^c_A2} as
\begin{equation}\label{eq:ScA3}
S^c_A={1 \over 1-n}\ln\left[{1 \over {\cal Z}^n(\beta)}\sum _d Z_A(\beta,d)Z_B^n(\beta,d)\right],
\end{equation}
where ${Z}_A(\beta,d)$ and ${ Z}_B(\beta,d)$ represent partition functions with boundary configuration $d$ for the two subsystems.

We can similarly write $S^q_A$ as
\begin{equation}\label{eq:SqA1}
S^q_A={1\over 1-n} \ln\left[ {1\over {\cal Z}^n(\beta)}\sum _{{\cal A}_i{\cal B}_i} e^{-{\beta\over 2}\left(E_{{\cal A}_1{\cal B}_1}+E_{{\cal A}_2{\cal B}_1}+E_{{\cal A}_2{\cal B}_2}\cdots E_{{\cal A}_1{\cal B}_n}\right)}\right],
\end{equation}
which yields a generalized version of Eq.~\eqref{eq:S^q}, i.e., $S^c_A={1 \over 1-n}\ln\left[{1 \over {\cal Z}^n(\beta)}\sum _d Z_A(\beta,d)^nZ_B^n(\beta,d)\right]$ (see Ref.~\onlinecite{Stephan2009} for closely related derivation based on Schmidt decomposition). Note that the energy of each replica subsystem ${\cal A}_i$ and ${\cal B}_i$ appears twice canceling the factor of $1\over 2$ in ${\beta \over 2}$.
 If the weights are ultralocal, we can then write
\begin{equation}
E_{{\cal A}{\cal B}_i}=\epsilon^{{\cal A}{\cal B}_i}_{a_1 a_2}+\epsilon^{{\cal A}{\cal B}_i}_{a_2 a_3}+\cdots \epsilon^{{\cal A}{\cal B}_i}_{a_h d}+\epsilon^{{\cal A}{\cal B}_i}_{d b_1^i}+\cdots \epsilon^{{\cal A}{\cal B}_i}_{b_{h-1}^i b_h^i},
\end{equation} 
where $a_i$ and $b_i$ represent configurations of rows parallel to the boundary. We then see that the weights can be absorbed into the transfer matrix $T$, and all steps of our derivation go through.

\bibliography{ice}{}
\end{document}